\documentclass[twocolumn,preprintnumbers,amsmath,amssymb]{revtex4}

\usepackage{graphicx}
\usepackage{dcolumn}
\usepackage{bm}
\usepackage{epstopdf}
\usepackage{xcolor}
\usepackage{float}

\begin{document}

\title{Effects of Laser Polarization and Color on Hybrid Laser Wakefield and Direct Acceleration}

\author{Xi Zhang$^{1,2}$, V. Khudik$^{1,2}$, A. C. Bernstein$^1$, M. C. Downer$^1$ and G. Shvets$^{1,2}$}
\affiliation{$^1$Department of Physics and Institute for Fusion Studies, The University of Texas at Austin, Austin, Texas 78712, USA}
\affiliation{$^2$School of Applied and Engineering Physics, Cornell University, Ithaca, New York 14850, USA}
\date{\today}

\begin{abstract}
We demonstrate that hybrid laser wakefield and direct acceleration (LWDA) can be significantly improved by using two laser pulses with different polarizations or frequencies. The improvement entails higher energy and charge of the accelerated electrons, as well as weaker sensitivity to the time delay between the leading (pump) pulse responsible for generating the wakefield, and the trailing laser pulse responsible for direct laser acceleration (DLA). When the two laser pulses have the same frequency, it is shown that it is advantageous to have their polarization states orthogonal to each other. A specific scenario of a frequency-doubled ($\lambda_{\rm DLA}=\lambda_{\rm pump}/2$) DLA laser pulse is also considered, and found to be an improvement over the single-frequency ($\lambda_{\rm DLA} = \lambda_{\rm pump}$) scenario.
\end{abstract}

\maketitle

\section{Introduction}
From their humble beginnings as a speculative theoretical idea~\cite{dawson_prl_lwfa}, laser-plasma acceleration has developed into one of the most exciting advanced accelerator techniques that has already lead to the production of monoenergetic electron beams~\cite{faure_nature04,geddes_nature04,mangles_nature04}, and is now pushing the energy frontier to multi-GeVs scale~\cite{leemans_nat,downer_nat_comm,kim_prl,leemans_prl} and beyond~\cite{steinke_nature}. Laser-wakefield acceleration (LWFA) is presently the most mature mechanism in laser-plasma electron acceleration. Rapid advances in laser technology enabled high-power high-intensity short pulse lasers capable of accessing the so-called full plasma blow-out ("bubble") regime~\cite{pukhov_bubble,li_prl}. Plasma electrons are fully blown out from the path of the laser pulse by its ponderomotive force $F_{p}\sim m_ec^2\nabla a^2$ (where $a=eA_L/mc^2$ is the normalized vector potential), thereby creating the plasma bubble in its wake. The plasma bubble regime is crucial for generating the high energy monoenergetic electron beams~\cite{faure_nature04,geddes_nature04,mangles_nature04}.

At the same time, other laser-plasma acceleration techniques have also been critically evaluated. One of the most promising among those techniques is the direct laser acceleration (DLA) approach, in which the laser electromagnetic field directly imparts energy to the electrons~\cite{pukhov_dla,gahn_prl}. DLA relies on the betatron resonance that occurs when the Doppler-shifted laser frequency $\langle \omega_d \rangle \equiv \omega_L(1-\langle v_x \rangle/v_{\rm ph})$ matches the $l'$th harmonic of the electron's betatron frequency $\omega_{\beta} = \omega_p/\sqrt{2 \langle \gamma \rangle}$: $\langle \omega_d \rangle = l\omega_{\beta}$, where $l=1,3,5 \ldots$ is an odd number that corresponds to either fundamental ($l=1$) or higher-order ($l>1$) betatron resonance~\cite{khudik1,khudik2}. Here $\langle v_x \rangle$ and $\langle \gamma \rangle$ are the time-averaged (over a betatron period) longitudinal velocity and relativistic factor of the accelerated electron; $\omega_L$ and $v_{\rm ph}$ are the frequency and the phase velocity of the laser field; $\omega_p = \sqrt{4\pi n e^2/m}$ is the electron plasma frequency.

Recently, the merger of the LWFA in the bubble regime and the DLA, which we refer to as the laser wakefield and direct acceleration (LWDA) mechanism, started attracting considerable
attention because of the possibilities for producing high energy electron beams for high energy physics, as well as copious high-energy X-rays~\cite{cary_prl,cipiccia,suk_pop,malka_prl11,phuoc_beta,phuoc_pop,mori_ppcf,ourprl,ourppcf} for a variety of applications that require high-brightness radiation. However, effective LWDA is not straightforward to achieve. Specifically, two requirements have been shown to be crucial for effective LWDA: (I) the initial energy $\epsilon_{\perp}(t=0)$ of betatron motion of an injected plasma electron must be sufficiently high to overcome its rapid reduction due to electron acceleration by the longitudinal field of the plasma bubble, (II) considerable overlap between the laser field and the injected electrons~\cite{ourprl,ourppcf}. The first requirement is fulfilled by either density bump injection~\cite{ourprl} or ionization injection~\cite{ourppcf,mori_ppcf}. Both approaches can deliver an injected electron beam with the initial transverse energy sufficient for effective DLA. 

The second requirement is met by time-delaying the second (DLA) laser pulse from the bubble-forming (pump) laser pulse~\cite{ourprl,ourppcf}. 
This requirement is exacerbated for moderate-power laser pulses (i.e. in the $10$TW range) which is also widely used~\cite{steinke_nature, li_prl,tsi_pop,malka_pre,kei_pop} because the relatively small size of the plasma bubble can result in undesirable interference between the two equal-frequency laser pulses. 
Such interference has two consequences. First, the shape of the plasma bubble can be considerably modified by the beatwave between the two pulses. As will be shown below, such modification can have a major effect on the total electron charge trapped inside the plasma bubble and subsequently accelerated by the combination of the LWFA and DLA mechanisms. Second, the peak laser field experienced by the injected electrons is also strongly modified by interference. This would naturally influence the electrons' energy gain from the DLA mechanism, as well as the fraction of the injected electrons that experiences the DLA. The latter point will also be elaborated on below. 

The overall conclusion that we draw from the above arguments is that both the charge and the energy distribution of the accelerated electrons could become highly sensitive to the time delay $\Delta t$ between the two pulses. Therefore, eliminating the interference between them could make the LWDA scheme considerably more robust. 

There are two natural approaches to eliminating interference. One is to preserve the equal wavelengths of the two laser pulses ($\lambda_{\rm DLA} = \lambda_{\rm pump}$), but to make their polarization states mutually orthogonal. 
The orthogonal polarization LWDA is promising to benefit both pitfalls. The other approach is used two-color LWDA with different wavelength of DLA and pump pulses. 
%

The effectiveness of the combination of the low and high frequency pulse has been demonstrated in ionization injection~\cite{xu_prstab,yu_prl,schoeder_prstab}.
The low frequency laser pulse with high ponderomotive force $F_p$ but low electric field $E_L$ blows out the plasma bubble and
the high frequency laser pulse with the opposite feature provides the ionization injection without disturbing the plasma bubble because the ionizing electric field in this pulse can be quite strong  $E_L\sim a/\lambda_L$ at small $\lambda$ and moderate $a$.

In this paper, we are exploring the similar combination  of frequencies of two pulses in the LWDA but using them the different purposes. The $\lambda_{\rm pump}=0.8\mu m$ low frequency laser pulse blows out the plasma bubble and produces
 the ionization of the high Z gas. The $\lambda_{DLA}=0.4\mu m$ second time-delayed high frequency laser pulse serves as the DLA pulse. It can potentially
increase the energy gain of electrons and enhance  the charge yield without disturbing the plasma bubble.  We also demonstrate the improved stability of the LWDA  mechanism by when high frequency   DLA pulse is used.

The remainder of the paper is organized as follows. In Sec.~\ref{sec:single} we carry out the simplified single-particle simulations and use them to address the basic questions such as the dependence of the number of DLA electrons on laser amplitude and the emittance growth of DLA electrons. The results of self-consistent 2D PIC simulations carried out using the VLPL code are presented in Sec.~\ref{sec:pic}.
Three scenarios are presented: single color parallel polarization LWDA (SP-LWDA), single color orthogonal polarization LWDA (SO-LWDA) and two color parallel polarization LWDA (TP-LWDA). We also discuss the potential application of the DLA electron beam as the radiation source. Conclusion and the directions for future research are outlined in Sec.~\ref{sec:conclusion}.

\section{Single-particle simulations}\label{sec:single}
Before moving into the self-consistent PIC simulations, we first illustrate the key requirements and the above mentioned basic questions by our simplified single-particle simulations.
The resulting equations of motion are given by \cite{cary_prl,phuoc_pop,phuoc_beta,ourprl,ourppcf}
\begin{eqnarray}
  \frac{dp_x}{dt} &=& -e \left( W_{x} - \frac{v_z}{c} B_y^{(L)} \right) \nonumber \\
  \frac{dp_z}{dt} &=&  -e \left( W_{z} + E_z^{(L)} + \frac{v_x}{c} B_y^{(L)} \right),\label{eq:model}
\end{eqnarray}

\begin{figure}[t]
\centering
   \includegraphics[height=0.36\textheight,width=0.98\columnwidth]{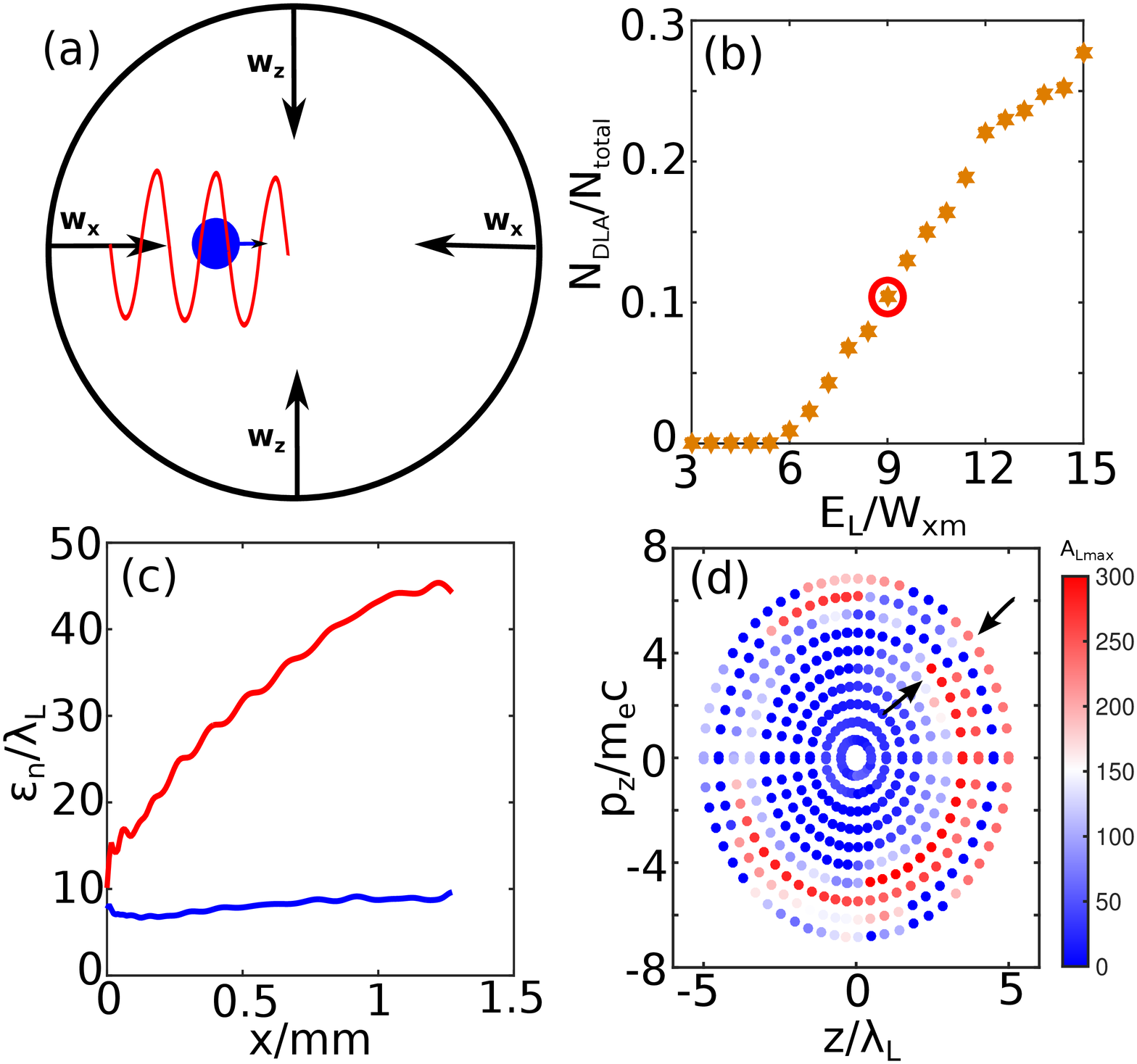}
\caption{Single-particle dynamics governed by Eqs.(\ref{eq:model}). (a) Schematic representation of single-particle simulation model. (b) The ratio between the number of DLA electrons and the number of trapped electrons. Empirical criteria for DLA electrons is $A_L/m_ec^2>0.3\gamma$. The red circle corresponds to the conditions in (c) and (d). (c) Evolution of the normalized emittance $\epsilon_n$ for both DLA electrons (red line) and non-DLA electrons (blue line). $ \epsilon_n = \langle p_x/m_ec \rangle \sqrt {\langle z^2 \rangle \langle (p_z/p_x)^2 \rangle - {\langle z p_z/p_x \rangle}^2} $, $\langle \rangle$ is the ensemble average over electrons. (d) Color-coded energy gain $A_L$ from the $\lambda_{0}=0.8\mu$m laser plotted as a function of the initial conditions with bubble accelerating field. Rings: $\epsilon_{\perp} ={\rm const}$.Simulation parameters for plots: $E_0 \approx 1.5 m_e c\omega_L/e$, $\gamma_{ph} = 14.4$, and $\omega_p/\omega_L = 0.093$, $E_0$ is changing in plot (b).}\label{fig:group}
\end{figure}

The schematic representation of our single-particle model is shown in Fig.~\ref{fig:group}. Electrons move under the influence of the bubble focusing field $W_{z}$, the bubble accelerating field $W_{x}$ and the laser fields $E_z$ and $B_y$. For simplicity, we assume planar linearly polarized laser fields in the form of $E_z^{(L)} = -E_0 \sin{\omega_L (t - x/v_{\rm ph})}$ and $B_y^{(L)} = B_0 \sin{\omega_L (t - x/v_{\rm ph})}$, where $B_0 = c E_0/v_{\rm ph}$ and $\omega_L$ is the DLA pulse frequency. The plasma bubble moves in the $x$-direction with velocity $v_b$ so that the bubble accelerating field is assumed to be $W_{x} = m\omega_p^2 (x-r_b-v_bt)/2e$, where the  bubble  radius $r_b=6\lambda_L$, $v_b=c(1-1/\gamma_b^2)^{1/2}$ and  $\gamma_b=10$.  The focusing fields inside the bubble can be approximated for relativistically moving electrons as $W_{z} = m\omega_p^2 z/2e$~\cite{kostyukov_pop04}.

One of the basic questions is the number of DLA electrons when the ratio between the maximum bubble field and the laser field $E_L/W_{xm}$ is changing. This
 is important since it is related to the charge yield of the LWDA. We use our single particle simulations to investigate the relation of the ratio between the number
 of DLA electrons and the total number of trapped electrons with the $E_L/W_{xm}$. Electrons are placed near the tail of the bubble and assigned a
 constant longitudinal momentum $p_x\sim\gamma_b mc$. The amplitude of DLA pulse $E_L$ is changing. As shown in Fig.~\ref{fig:group}(b), $N_{DLA}/N_{total}$ is increasing with respect to $E_L/W_{xm}$. Therefore, the large laser field is good to generate more DLA electrons. The high frequency DLA pulse that has larger $E_L$ is able to increase the energy gain from laser since $A_L \propto \int E_L v_{\perp}dt$ and also the charge yield.

The other basic question of LWDA is the emittance of the electron beam. Usual ionization injection produces electrons with large initial emittance\cite{xu_prstab}. It is not good for LWFA but LWDA can utilize it. In our single particle simulations, the initial emittance is about $\epsilon_n \simeq 6 mm\cdotp mrad$. The emittance for non-DLA electrons does not change a lot during the whole acceleration process but the DLA electrons have about 6 times larger normalized emittance $\epsilon_n$ than the initial emittance and  continuously grows during the acceleration process as shown in Fig.~\ref{fig:group}(c).

We look into the case marked by the red circle in Fig.~\ref{fig:group}(b) in detail and show the characters in Fig.~\ref{fig:group}(d). The initial conditions and the simulation parameters in the caption of Fig.~\ref{fig:group} were consistent with PIC simulations presented in Sec.~\ref{sec:pic}. Initially (at $t=x=0$ ) electrons are assigned a constant longitudinal momentum $p_x\sim\gamma_b mc$ that is the prerequisite for trapping electrons in the plasma bubble. The transverse initial conditions $(z_0,p_{z0})$ are chosen randomly inside the $0 < \epsilon_{\perp} \leq 2.1 m_ec^2$ phase space ring as shown in Fig.~\ref{fig:group}(d). The transverse energy is defined $\epsilon_{\perp} = p_{\perp}^2/2\gamma m_e + m_e \omega_{p}^2 z^2/4$~\cite{ourprl,ourppcf,kostyukov_pop04,malka_prl11}. We also assume that the electrons overlap with the DLA pulse all the time in the simplified model.

A LWDA relies on the large initial transverse energy $\epsilon_{\perp}(t=0)$~\cite{ourprl,ourppcf,kostyukov_pop04,malka_prl11} and the spatio-temporal overlap between injected electrons and the laser field. It has been demonstrated that the density bump injection~\cite{ourprl} and the ionization injection~\cite{ourppcf,mori_ppcf} are able to produce large enough $\epsilon_{\perp}(t=0)$. We analyze the case that is indicated by the red circle in Fig.~\ref{fig:group}(b) and plot the initial phase space of electrons color-coded by the energy gain from laser in Fig.~\ref{fig:group}(d). The electrons have to have large transverse energy $\epsilon_n(t=0)$ to get significant DLA and the DLA electrons ($A_L>200m_ec^2$) occupy a wide band with $1.0 m_ec^2 < \epsilon_{\perp}(t=0) < 2.1 m_ec^2$. This is good for generating more DLA electrons. To understand this effect, we take a look at The ultra-relativistic limits of $\langle\omega_d\rangle$ and $\omega_{\beta}$  which has the expressions as follows:
\begin{equation}\label{eq:freq}
  \langle \omega_d \rangle \simeq \omega_L \left( \frac{1 + \langle p_z^{2} \rangle /m_e^{2} c^2}{2\gamma^2} + \frac{1}{2\gamma_{\rm ph}^{2}} \right), \ \ \ \omega_{\beta} \simeq \frac{\omega_p}{\sqrt{2\gamma}},
\end{equation}
where $\gamma_{\rm ph} \equiv 1/\sqrt{v_{\rm ph}^{2} - 1}$. The betatron resonance requires $\langle \omega_d \rangle=\omega_{\beta}$. It is obvious that the large $\epsilon_n(t=0)$ is prerequisite based on Eq.\ref{eq:freq}. But the large $\epsilon_n(t=0)$ has wide choice because the longitudinal accelerating field introduces one more degree of freedom that can increase $\gamma$ but without raising $p_z$. The requirement of $\epsilon_{\perp}(t=0)$ is relaxed due to the longitudinal accelerating field and more electrons with large $\epsilon_n(t=0)$ have the chance to undergo DLA in the bubble regime.

\section{Particle-in-Cell Simulations}\label{sec:pic}

\begin{figure}[ht]
\begin{center}
\vspace{2 mm}
\includegraphics[height=0.39\textheight,width=0.98\columnwidth]{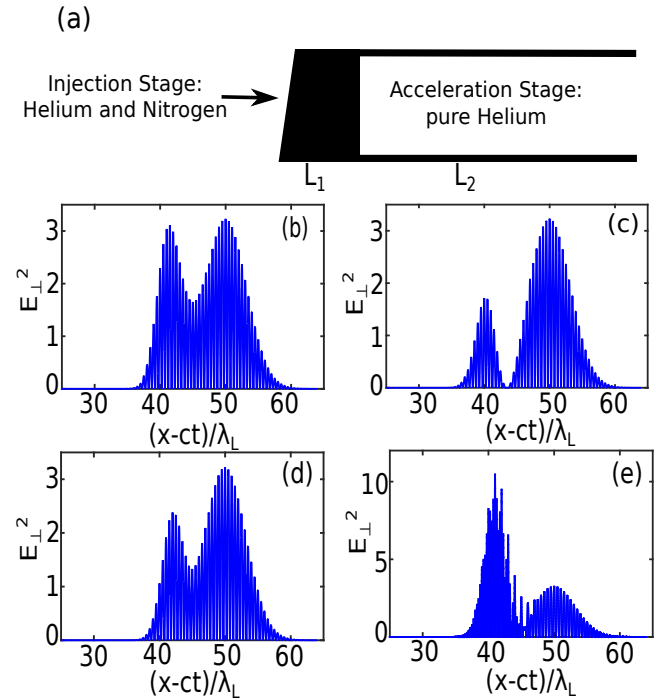}
\end{center}
\caption{(a) plasma density profile. It is divided into the injection stage and acceleration stage. (b) Initial on axis $E_{\perp}^2$ for the single color parallel polarization case $\lambda_{\rm pump}=\lambda_{DLA}=\lambda_L$ and time delay $\Delta \tau = 24fs$. (c) Initial on axis $E_{\perp}^2$ for the single color parallel polarization case $\lambda_{\rm pump}=\lambda_{DLA}=\lambda_L$ and time delay $\Delta \tau = 25.3fs$, the destructive interference appears.(d) Initial on axis $E_{\perp}^2$ for the single color orthogonal polarization case $\lambda_{\rm pump}=\lambda_{DLA}=\lambda_L$ and time delay $\Delta \tau = 21fs$. (e) Initial on axis $E_{\perp}^2$ for the two color parallel polarization case $\lambda_{\rm pump}=2\lambda_{DLA}=\lambda_L$ time delay $\Delta \tau = 24fs$. $E_{\perp}$ is normalized to $m_ec\omega_L/e$. Plasma parameters: mixed gas length $L_1 = 100\mu m$, $L_2 \simeq 1$mm; $n_0 = 1.5\times 10^{19}cm^{-3}, n_{N^{5+}} = 0.2n_0$, ionization potential for $N^{5+}\rightarrow N^{6+}$ $U_{ion} \simeq 552.1ev$ and $N^{6+}\rightarrow N^{7+}$ $U_{ion} \simeq 667.0ev$, $\lambda_p = 2\pi c/\omega_p = 9\mu m$. Laser parameters: wavelength $\lambda_{\rm pump} = \lambda_L$, $I_{\rm pump} = 7.0\times 10^{18}$W/cm$^2$, $I_{\rm DLA} = 4.8\times 10^{18}$W/cm$^2$ for $\lambda_{DLA}=\lambda_L$ and $I_{\rm DLA} = 1.9\times 10^{19}$W/cm$^2$ for $\lambda_{DLA}=0.5\lambda_L$, pulse durations $\tau_{\rm pump} =20$fs and $\tau_{\rm DLA} = 10$fs, spot size $w_{\rm pump} = 10\mu m$, $w_{\rm DLA} = 10.6\mu m$ for $\lambda_{DLA}=\lambda_L$ DLA pulse and $w_{\rm DLA} = 5.3\mu m$ for $\lambda_{DLA}=0.5\lambda_L$ DLA, inter-pulse time delay is changing $\Delta \tau = 21.3 - 26.7$fs. Simulation parameters: numerical grid’s cell size $\Delta x \times \Delta z = \lambda_L/50 \times \lambda_p/50$ for $\lambda_{DLA}=\lambda_L$ DLA pulse and $\Delta x \times \Delta z = \lambda_L/100 \times \lambda_p/50$ for $\lambda_{DLA}=0.5\lambda_L$ DLA pulse. $\lambda_L=0.8\mu m$}\label{fig:densprof}
\end{figure}

In this Section, we use first-principles self-consistent relativistic 2D PIC code VLPL~\cite{pukhov_vlpl} to simulate the effects of laser polarization and color on LWDA. The schematic of a proposed LWDA is shown
in Fig.~\ref{fig:densprof} (see caption for simulations parameters). We consider the scheme that is composed of the short length injection stage and acceleration
stage. This two-stage scheme has been experimentally verified~\cite{pollok_prl,vargas_apl} and the sub-hundred micrometer gas nozzel has also been achieved~\cite{schmid_rsi,jolly_rsi}. Ten-terawatt pump pulse ($P_{\rm pump}=12$TW) and the time-delayed DLA pulse ($P_{\rm DLA} = 10$TW) are assumed in the simulations. The laser
parameters are consistent with the UT$^3$ laser system~\cite{li_prl,tsi_pop}. We keep the power and $a_0$ the same when the $\lambda_{DLA}=\lambda_L/2$ DLA pulse is used. It means that the $\lambda_{DLA}=\lambda_L/2$ DLA pulse is tightly focused. The $100 \mu m$-long injection stage is formed by gas mixture of $80\%$ $H_e$ and $20\%$ $N_{2}$ shown as a dark area in the middle of Fig.~\ref{fig:densprof}. The injection is due to the ionization of high-Z nitrogen ions ($N^{5+}$ and $N^{6+}$) in the simulations. We
have demonstrated that the ionization injection is able to produce electrons with large enough $\epsilon_{\perp}(t=0)$ for the operation of the LWDA~\cite{ourppcf}.

\begin{figure}[h]
\centering
   \includegraphics[height=0.32\textheight,width=0.9\columnwidth]{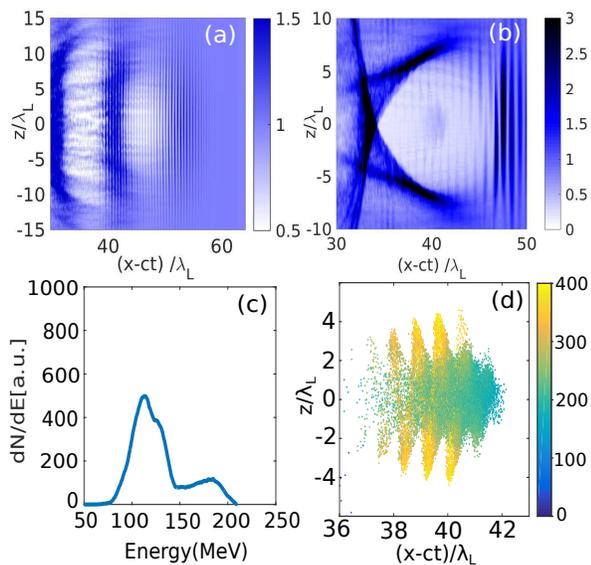}
\caption{(a) Nonlinear wake structure at the propapation distance $x=128\mu$m for the SP-LWDA. (b) The bubble structure at $x=640\mu$m for the SP-LWDA. The density is normalized to the initial density in both (a) and (b). (c) the energy spectrum for electrons in (d). (d) Spatial distribution of the trapped electrons and color-coded by their $\gamma$ at $x=640\mu$m for the SP-LWDA.}\label{fig:potential}
\end{figure}

The background plasma is formed by the leading edge ionization of the neutral gas. The inner shell nitrogen's electrons ($N^{5+}\rightarrow N^{6+}$ and $N^{6+}\rightarrow N^{7+}$) are produced via ionization close to the peak of the pump pulse intensity and are injected and get trapped inside the plasma bubble~\cite{pak,liu_prl,mc_prl}. The ionizations also happen for the DLA pulses but due to the relative position of the DLA pulses, the wake potential difference can not satisfied the trapping condition $e\Delta \psi/m_ec^2\simeq -1$~\cite{pak,xu_prstab,schoeder_prstab,ourppcf}. Therefore, the pump pulse plays the roles of producing the plasma bubble and injecting the electrons into it. The ADK tunneling ionization model~\cite{adk,chen_jcp,pukhov_ion} is used to describe the release of the inner shell electrons from the nitrogen ions.

Figs.~\ref{fig:densprof}(b-d) are the initial on axis $E_{\perp}^2$ in three schemes we consider. Figs.\ref{fig:densprof}(b) and (c) are the single color case. The single color parallel polarization LWDA has larger initial on axis $E_{\perp}^2$ in the DLA pulse part than single color orthogonal polarization LWDA due to the constructive interference. The two color parallel polarization LWDA has the same $a_0$ as single color LWDAs so the initial on axis $E_{\perp}^2$ is much higher that is clear by looking at Fig.\ref{fig:densprof}(c). The optimal time delay of the scecond DLA pulses are $\Delta \tau = 24fs$ for the single color parallel polarization LWDA and the two color parallel LWDA and $\Delta \tau = 21fs$ for the single color orthogonal polarization LWDA regarding the peak DLA electron energy.

\subsection{Single Color Parallel Polarization LWDA (SP-LWDA)}\label{subsec:sp}
We first look into the details of the SP-LWDA. As shown in Fig.~\ref{fig:potential}(a), the plasma bubble has not formed during the injection stage. The trapping of electrons happens in the nonlinear wake stage. With the self-focusing of the pump pulse, the plasma bubble is formed at about $x=200\mu$m. Fig.~\ref{fig:potential}(b) shows the bubble structure at the end of the propagation. There are electrons that are accelerated inside of the plasma bubble but they are not dense. The zoom-in color-coded electron spatial distribution is shown in Fig.~\ref{fig:potential}(d). The DLA electrons advance slower than the non-DLA electrons. Detail discussions are in ~\cite{ourprl,ourppcf}. The spectrum is in Fig.~\ref{fig:potential}(c). Two peaks are formed. The DLA electrons peaks at about $\gamma^{\rm DLA} m_ec^2 \approx 170$MeV and gain about $60$MeV more energy than non-DLA electrons. The vertical axis in Fig.~\ref{fig:potential}(c) is the number of macro-electrons. It is for the later comparison with the other two cases. We observe all the features of LWDA as mentioned in~\cite{ourprl,ourppcf} but in this moderate power regime.

\begin{figure}[h]
\centering
   \includegraphics[height=0.64\textheight,width=0.9\columnwidth]{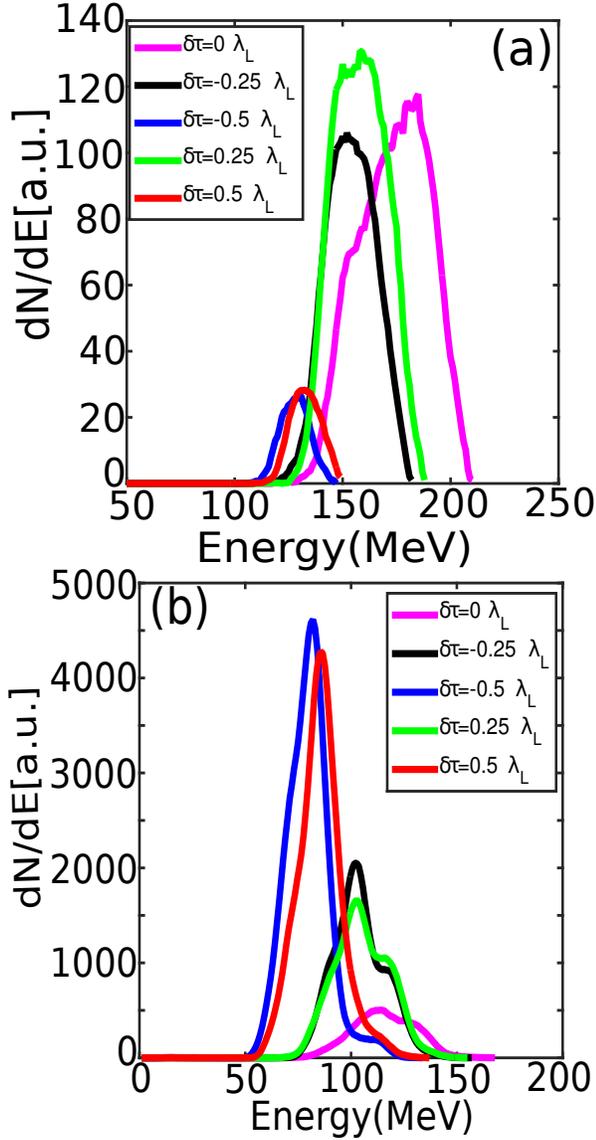}
\caption{(a) The spectra for the DLA electrons $A_L>=75 MeV$ in different cases. (b) The spectra for the non-DLA electrons $A_L<75 MeV$ in different cases}\label{fig:specs}
\end{figure}

The time-delay of the second DLA pulse is one of the crucial parameters. This parameter is from the synchronization of the pump pulse and the DLA pulse. In the
moderate laser power regime (e.g. UT$^3$ laser system), the plasma bubble is smaller $R\sim \sqrt{a_0}/k_p$~\cite{gordienko} than the cases in~\cite{ourprl,ourppcf}
and the interference between the pump pulse and the DLA pulse has to be taken into account. The SP-LWDA is the most common setup but it relies on the extremely
accurate timing of the time-delay $\Delta \tau$. To understand how the time-delayed of the DLA pulse affect the LWDA, we need to consider different cases that the
time delay of the DLA pulse is slightly shifted from the position in Figs.~\ref{fig:potential}. Fig4.~\ref{fig:specs} shows several cases with both forward shifted $\Delta \tau=0.25\lambda_L, 0.5\lambda_L$ and the backward shifted $\Delta \tau=-0.25\lambda_L, -0.5\lambda_L$ in the SP-LWDA.

We have already seen the on axis $E_{\perp}^2$ for the case $\delta\tau = -0.5\lambda_L, 0\lambda_L$ in Fig.\ref{fig:densprof}(b,c). Fig.~\ref{fig:densprof}(b) is the constructive interference and the destructive interference appears by only shifting the DLA pulse $0.5\lambda_L$ backward as shown in Fig.~\ref{fig:densprof}(c). The $E_{\perp}^2$ at the bottom of the bubble are changed significantly. These cases are only slight shifts but the SP-LWDAs produce the drasticly different results. Since the number of the DLA electrons may become very small, we roughly seperate the energy spectrum to the DLA spectrum and the non-DLA spectrum by the empirical criteria that the electrons gain $A_L>=75 MeV$ directly from the laser field $E_L$ belong to the DLA group. The DLA and non-DLA spectra for all five cases are shown in Fig.\ref{fig:specs}(c) and Fig.~\ref{fig:specs}(d) repectively. The magenta curves in Fig.~\ref{fig:specs}(a,b) are the DLA spectrum and the non-DLA spectrum for $\delta\tau = 0\lambda$, corresponding to $\Delta\tau = 24fs$. From the point of view of the DLA energy gain, it is clear that $\Delta\tau = 24fs$ is the optimal case. The peak energy of the DLA spectra changes from about $\gamma^{\rm DLA} m_ec^2 \approx 120$MeV to $\gamma^{\rm DLA} m_ec^2 \approx 170$MeV. The number of the accelerated macro-electrons has about 6 times different from the best case (the magenta curve) to the worst case ( the blue curve). Only sightly shifts of the $\lambda_{DLA}=\lambda_L$ DLA pulse deteriorate the final results. We observe that the number of DLA electrons increase in the constructive interference although the total number of the trapped electrons drop. This phenomenon can be explained by Fig.\ref{fig:group}~(b). The ratio of DLA electrons increases with the DLA pulse peak electric field and the constructive interference broadens the high laser electric field region.
{}
\begin{figure}[h]
\centering
   \includegraphics[height=0.18\textheight,width=0.95\columnwidth]{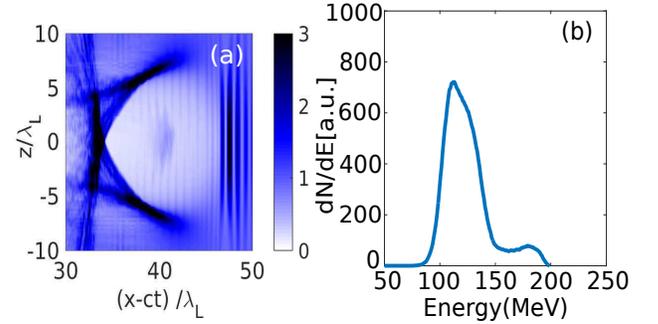}
\caption{(a) The bubble structure at $x=640\mu$m for the SO-LWDA. The density is normalized to the initial density. (b) the energy spectrum for electrons trapped in the bubble.}\label{fig:orth}
\end{figure}

Note that the non-DLA spectra in Fig.\ref{fig:specs}(b) also exhibit the differences in the peak energy and the number of the accelerated macro-electrons. The differences in the number of the accelerated macro-electrons are due to the perturbation of the laser wake induced by the $\lambda_{DLA}=\lambda_L$ DLA pulses in the injection stage. The differences in the peak energy are from the beam loading effect~\cite{ourprl,tsung_prl,tzoufras_prl} which influences the plasma bubble accelerating field. We have to mention that the differences in the DLA spectra are influenced by both the energy gain from laser and the energy gain from wake. It is convincing that the time delay of the $\lambda_{DLA}=\lambda_L$ DLA pulse has to be carefully selected and the SP-LWDA is not very stable in the moderate laser power regime. Therefore, we have to turn to the other directions to avoid the time jittering of LWDA.

\begin{figure}[h]
\begin{center}
\vspace{2 mm}
\includegraphics[height=0.60\textheight,width=0.9\columnwidth]{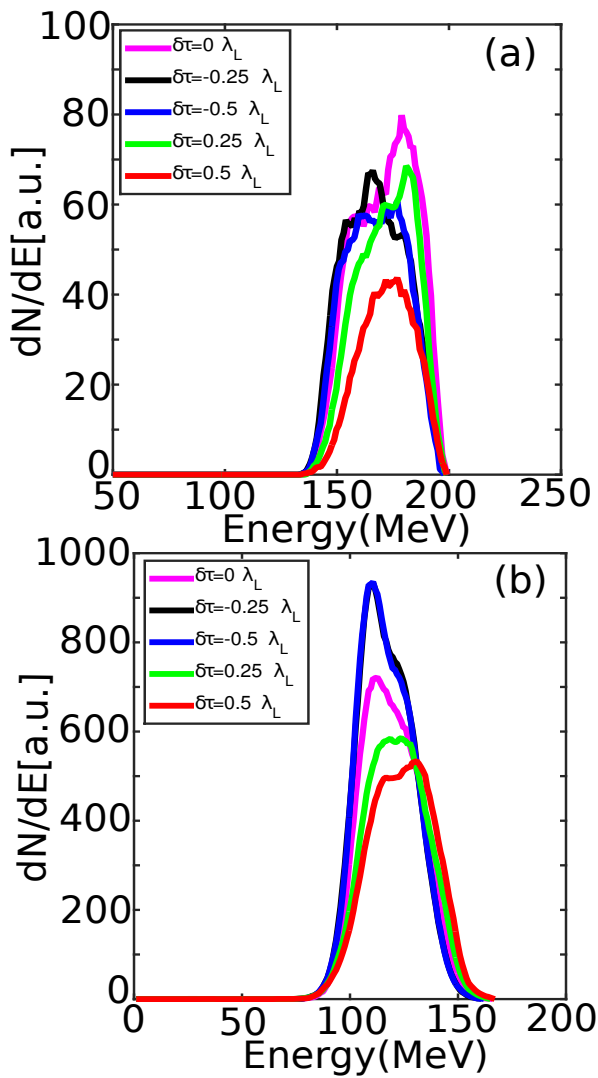}
\end{center}
\caption{Electron energy spectra in the SO-LWDA. (a) The energy spectra for the DLA electrons $A_L>=75 MeV$ in both forward shift and backward shift cases. (b) The energy spectra for the non-DLA electrons $A_L<75 MeV$ in both backward shift and forward shift cases.}\label{fig:orspecs}
\end{figure}

\subsection{Single Color Orthogonal Polarization LWDA (SO-LWDA)}\label{subsec:so}
The interference between the pump pulse and the DLA pulse is the main resource of the time jittering
of SP-LWDA. The natural way to avoid this time jittering is to rotate the pump pulse so that the DLA pulse and the pump pulse are
orthogonal to each other. We investigate this situation by making the pump pulse polarize in the Y direction. Other laser and plasma parameters are exactly the same
as shown in the SP-LWDA.

As shown in Fig.~\ref{fig:orth}(a), the plasma bubble structure does not have observable difference compared with Fig.~\ref{fig:potential}(b). But the numbers of 
trapped electrons are different between SP-LWDA and SO-LWDA. This leads to the different accelerating field due to the beam loading effect. We 
have to mention that the optimal time-delays of the DLA pulse have a small difference 
(see caption of Fig.~\ref{fig:densprof}) between the SP-LWDA and SO-LWDA because of the evolution of the DLA
pulse. The spectra in the optimal cases of the SP-LWDA and the SO-LWDA are in general similar. Non-DLA electrons peaks at about 
$\gamma^{\rm non-DLA} m_ec^2 \approx 110$MeV and DLA electrons peaks at about $\gamma^{\rm DLA} m_ec^2 \approx 170$MeV. The number of the macro-electrons are different. 
The difference in the trapped electrons is from the perturbation to the nonlinear wake in the injection stage. We will discuss it in detail later. There is also 
difference in the number of the DLA electrons. 
The high energy peak in the spectrum of Fig.~\ref{fig:potential}(c) is higher than in Fig.~\ref{fig:orth}(b) although the total number of the trapped electrons is 
smaller. The reasons are two-fold: one is that the constructive interference increases the DLA pulse electric field so the electrons experience stronger laser field in the optimal case of the SP-LWDA; the other is that there are more ricochet electrons in parallel polarization LWDA\cite{ourppcf}.

The time jittering is improved in the SO-LWDA. The variations in the DLA and the non-DLA spectra are smaller in
Fig.~\ref{fig:orspecs} than in Fig.~\ref{fig:specs}. Fig.~\ref{fig:orspecs}(a) are the DLA spectra for the both backward shift case
and the forward shift case. The variations in the peak energy and the number of electrons are better but there are still some fluctuations. Since there is no
interference between the pump pulse and the DLA pulse, the overlap between the DLA pulse and the trapped electrons plays the major role in the spectra fluctuations.
For non-DLA spectra as shown in Fig.~\ref{fig:orspecs}(b), the fluctuations in the spectra are from the perturbations to the nonlnear wake in the injection stage as
mentioned above. We have seen that the time-jitterings of the final spectra are getting better by rotating the pump pulse to orthogonal direction with
respect to the DLA pulse.

\begin{figure}[h]
\centering
   \includegraphics[height=0.18\textheight,width=0.95\columnwidth]{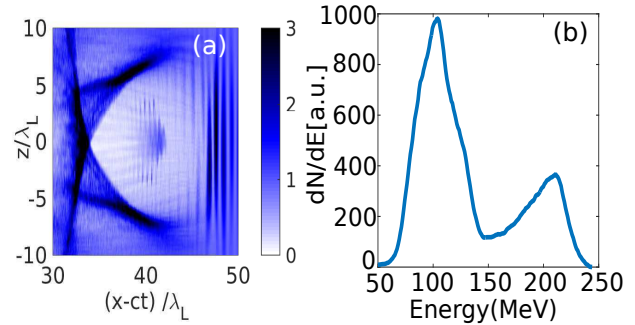}
\caption{(a) The bubble structure at $x=640\mu$m for the TP-LWDA. The density is normalized to the initial density. (b) the energy spectrum for electrons trapped in the bubble.}\label{fig:dual}
\end{figure}

\subsection{Two Color Parallel Polarization LWDA (TP-LWDA)}\label{subsec:tp}
By rotating the polarization direction of the pump pulse, we improve the performance of the LWDA. It is
possible that the performance can be further improved by introducing the $\lambda_{DLA}=0.5\lambda_L$ DLA pulse. The wake structure is mainly determined by the $a_0$. 
We keep the initial $a_0$ of the DLA pulse the same as in SP-LWDA and SO-LWDA. Since $a_0\propto \sqrt{I}\lambda_L$, the laser electric field can be higher without 
changing the wake structure. The energy gain from the laser is $A_L \propto \int E_L v_{\perp}dt$. So the electrons are able to gain higher energy from the laser in 
TP-LWDA. The other interesting point of the TP-LWDA is that the interference between the pump pulse and the DLA pulse does not have strong influence because of the 
different frequenies.

\begin{figure}[h]
\centering
   \includegraphics[height=0.18\textheight,width=0.95\columnwidth]{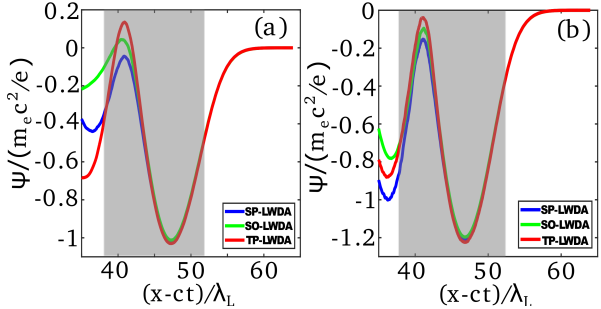}
\caption{(a) On axis wake potential $\psi$ at propagation distance $x=64\mu$m for SP-LWDA(blue), SO-LWDA(green) and TP-LWDA(red). (b) On axis wake potential $\psi$ at propagation distance $x=128\mu$m for SP-LWDA(blue), SO-LWDA(green) and TP-LWDA(red). More than $95\%$ of ionization happen in the gray region}\label{fig:onxpsi}
\end{figure}

The bubble structure is shown in Fig.~\ref{fig:dual}(a). It looks similar with Fig.\ref{fig:potential}(b) and in Fig.\ref{fig:orth}(a). The DLA pulses do not influence
the plasma bubble but they do have effects on the evolution of the nonlinear wake in the injection stage. The trapped electrons are denser in Fig.~\ref{fig:dual}(a).
Without the perturbation to the non-linear wake from the constructive interference, there are more electrons that are trapped. Note that the trapped electrons are 
smaller in in SO-LWDA than TP-LWDA although both of them do not have constructive interference. The reason is that the optimal position of the DLA pulse is shifted in 
SO-LWDA and it has the effects on the nonlinear wake evolution. To get a clear picture of the trapping dynamics in the injection stage, we look into the on axis wake 
potential $\psi$ as shown in Fig.\ref{fig:onxpsi}. Due to the pump pulse and the DLA pulse, more than $95\%$ of the ionization happens in the gray colored region as 
shown in Fig.\ref {fig:onxpsi}. The trapping condition is $e\Delta\psi/m_ec^2\simeq-1$~\cite{pak,chen_pop,ourppcf}. In Fig.\ref{fig:onxpsi}(a), the TP-LWDA (red 
curve) is the highest and close to $e\psi/m_ec^2 \simeq 0.2$ at the bottom part of the nonlinear wake. The trapping condition is satisfied for TP-LWDA (red curve) in 
a widest region. In the contrary, the trapping condition can not be satisfied for SP-LWDA (blue curve) at propagation distance $x=64\mu$m. As the nonlinear wake 
evolved, all three scenarios can satisfy the trapping condition as shown in Fig.\ref{fig:onxpsi}(b). The region for TP-LWDA is still the widest. Therefore, we 
understand that the charge yield for TP-LWDA is the highest.

Not only the number of the trapped electrons, the energy of the DLA electrons are also higher in TP-LWDA. As shown in Fig.\ref{fig:dual}, the DLA peak is at about
$\gamma^{\rm DLA} m_ec^2 \approx 210$MeV. It almost doubles the energy peak of the non-DLA electrons and is close to $50$MeV higher than SP-LWDA and SO-LWDA without
brodening the beam's energy spread. The other interesting effect is the number of the DLA electrons. The number of DLA electrons in optimal TP-LWDA counts for about
$35\%$ of the number of the non-DLA electrons while this ratio is about $20\%$ for optimal SP-LWDA and about $12\%$ for optimal SO-LWDA. The increase of the total
number of trapped electrons is one reason. The other reason is that the laser intensity is higher in TP-LWDA, which has the connection to the number of the DLA
electrons as indicated in Fig.\ref{fig:group}(b).

\begin{figure}[h]
\begin{center}
\vspace{2 mm}
\includegraphics[height=0.60\textheight,width=0.9\columnwidth]{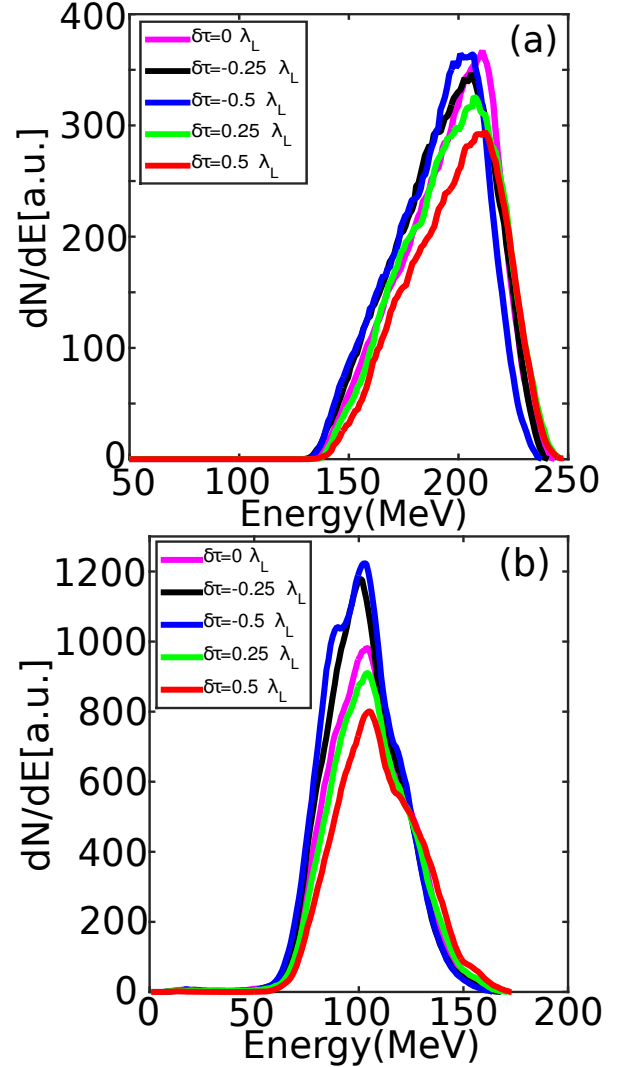}
\end{center}
\caption{Electron energy spectra in TP-LWDA. (a) The energy spectra for the DLA electrons $A_L>=75 MeV$ in both backward shift and forward shift cases. (b) The energy spectra for the non-DLA electrons $A_L<75 MeV$ in both backward shift and forward shift cases.}\label{fig:color}
\end{figure}

Higher energy gain and higher charge yield are first two-folds advantages. The stability is also important. A few cases with the slightly different time delay of the $
\lambda_{DLA}=0.5\lambda_L$ DLA pulse are considered. It is boring to see the more or less the same plots as in Fig.\ref{fig:densprof}(d) for both the backward shift 
case and the forward shift case. So we directly show the final spectra for the DLA electrons and the non-DLA electrons. The rough seperation criteria is still the 
electrons gain $A_L>=75 MeV$ directly from the laser field. On the side of the final electron energy, no matter how to shift the $\lambda_{DLA}=0.5\lambda_L$ DLA 
pulse, the DLA spectra consistently peak at about $\gamma^{DLA}m_ec^2 = 200 MeV \sim 210 MeV$ as shown in Fig.\ref{fig:color}(a). On the other side of the number of the accelerated 
macro-electrons, there are fluctuations but not as drastic as in the single color cases. The worst case is $\delta\tau = 0.5\lambda_L$ forward from the initial time 
delay $\Delta\tau = 24fs$ as indicated by the red line in Fig.\ref{fig:color}(a). The number of the DLA electrons is still about two times higher than the optimal 
case in the SP-LWDA in Fig.\ref{fig:specs}(a) and three times higher than the optimal case in SO-LWDA as shown by the magenta curve in Fig.\ref{fig:orspecs}(a). The 
ratio between the number of the DLA electrons to the non-DLA electrons are between 20$\%$ to 35$\%$ for all the cases. This is significantly better than the most of 
cases in the SP-LWDA and in the SO-LWDA in which the ratios drasticaly change from 20$\%$ to about 1$\%$. We find that the non-DLA spectra in the forward shift cases 
and the backward shift cases are also slightly different in the number of the trapped electrons and the peak energy. The explanation for this phenomenon is similar 
with the SO-LWDA. It is due to the perturbation of the laser wake induced by the position variation of the $\lambda_{DLA}=0.5\lambda_L$ DLA pulse. From the 
investigation of the TP-LWDA, the high frequency DLA pulse is optimistic to the stable operation of the LWDA.

\begin{figure}[h]
\centering
   \includegraphics[height=0.32\textheight,width=0.9\columnwidth]{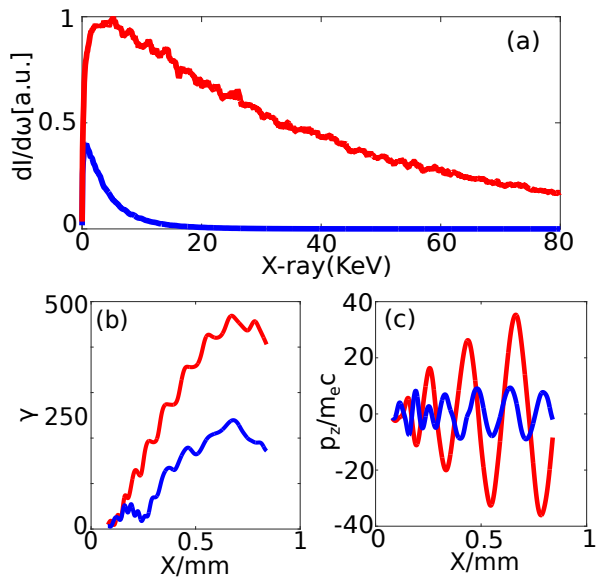}
\caption{(a) X-ray spectra for the representative DLA electron (red) and non-DLA electron (blue) in TP-LWDA. (b) the evolution of $\gamma$ for the representative DLA electron (red) and non-DLA electron (blue). (c) the transverse momentum $p_z$ for the representative DLA electron (red) and non-DLA electron (blue).}\label{fig:xray}
\end{figure}

The DLA electrons are able to produce more copious X-rays than the non-DLA electrons. We select two representative electrons from the simulation of TP-LWDA. The DLA electron has larger $p_z$ and $\gamma$ than non-DLA electron as shown in Fig.~\ref{fig:xray}(b) and Fig.~\ref{fig:xray}(c). The X-ray spectra are calculated by integrating over the electrons' trajectories based on the synchrotron radiation formula in~\cite{jackson}. As shown in Fig.~\ref{fig:xray}(a), the DLA electron has much higher and wider X-ray spectrum than the non-DLA electron. The DLA electron has $\gamma_{max}\simeq 450 m_ec^2$ and $p_{z}^{max}\simeq 35 m_ec$ and the non-DLA electron has $\gamma_{max}\simeq 250 m_ec^2$ and $p_{z}^{max}\simeq 8 m_ec$. We can estimate the maximum X-ray critical frequency $\omega_c = 1.5 \gamma^3 c/\rho$~\cite{jackson} for the DLA electron is $\omega_c^{DLA} \sim 45$KeV and for the non-DLA electron is $\omega_c^{nDLA} \sim 5$KeV. The DLA and non-DLA X-ray spectra roughly peak at about $\omega_{peak}^{DLA} \sim 7$KeV and $\omega_{peak}^{nDLA} \sim 0.6$KeV respectively. Therefore, the DLA electrons could be the excellent radiation source.

\section{Conclusion}\label{sec:conclusion}
In conclusion, we have investigated the three possible scenarios of laser wakefield and direct acceleration (LWDA) in a moderate power regime: SP-LWDA, SO-LWDA and TP-
LWDA. SP-LWDA has huge time-jittering in the final electron spectrum because of the interference between the pump pulse and the DLA pulse. SO-LWDA has relatively 
better performance than the SP-LWDA since it eliminats the interference and improves the time jittering. But the DLA charge yield is not improved. TP-LWDA combines 
the benefits of the $\lambda_{\rm pump}=\lambda_L$ pump pulse and the $\lambda_{DLA}=0.5\lambda_L$ DLA pulse. It is demonstrated that TP-LWDA achieves higher energy 
and higher charge DLA electrons compared with the SP-LWDA and SO-LWDA. Furthermore, the TP-LWDA increases the stability by lowering the requirement of synchronization 
between the pump pulse and the DLA pulse. With the introduction of the frequency upshift DLA pulse, the hybrid laser wakefield and direct laser plasma accelerator may 
become even more valuable for future experiment realization.

This work was supported by DOE grants DE\-SC0007889  and by an AFOSR grant FA9550-16-1-0013. The authors thank the Texas Advanced Computing Center for providing HPC resources.

\end{document}